\documentclass[twocolumn,         % Format : preprint, twocolumn
               showpacs,            % Pacs : showpacs, noshowpacs
               preprintnumbers,     % Preprint: preprintnumbers,
               aps,                 % Society: ...
               prd,          	    % Journal Style : pra, prb, prc, prd, pre,
               a4paper,             % Size : a4paper, ...
               superscriptaddress,      % Affiliation (Title) : groupedaddress,
               nofootinbib,         % Footnote: footinbib, nofootinbib
               tightenlines,        % Remove additional spaces in a line
               floats,floatfix,11pt
               ]{revtex4-2}              
\usepackage{graphicx}  % needed for figures
\usepackage[a4paper, total={7.0in, 9.5in}]{geometry}
\usepackage{dcolumn}   % needed for some tables
\usepackage{bm}  
\usepackage[normalem]{ulem}
\usepackage{subcaption}

\usepackage[normalem]{ulem}
\usepackage[utf8]{inputenc}
\usepackage{amsmath,amssymb}
\usepackage{soul}
\usepackage{float}
\usepackage{makecell}
\usepackage[thinlines]{easytable}
\usepackage{array,booktabs}
\usepackage{lipsum} 
\usepackage[colorlinks=true,linkcolor=blue,citecolor=blue]{hyperref}
\usepackage{subcaption}
\usepackage{array,makecell}
\begin{document} 

\title{Quintom Dark Energy: Future Attractor and  Phantom Crossing in Light of DESI DR2 Observation}
\author{Phusuda Thanankullaphong }
\email{phusuda.tha@student.mahidol.ac.th}
\affiliation{NAS, Centre for Theoretical Physics \& Natural Philosophy, Mahidol University,
Nakhonsawan Campus, Phayuha Khiri, Nakhonsawan 60130, Thailand}

\author{Prasanta Sahoo }
\email{prasantmath123@yahoo.com} 
\affiliation{Midnapore College (Autonomous), Midnapore, West Bengal, India, 721101}
\affiliation{NAS, Centre for Theoretical Physics \& Natural Philosophy, Mahidol University,
Nakhonsawan Campus, Phayuha Khiri, Nakhonsawan 60130, Thailand}

\author{Prajwal Hassan Puttasiddappa}
\email{prajwal.puttasiddappa@edu.ufes.br}

\affiliation{PPGCosmo, Universidade Federal do Esp\'irito Santo, 29075-910, Vit\'oria, ES, Brazil}
\affiliation{Departamento de F\'isica Te\'orica, Instituto de F\'isica, Universidade do Estado do Rio de Janeiro (UERJ), Rua Sao Francisco Xavier 524, Maracana CEP 20550-013, Rio de Janeiro, RJ, Brazil}
\affiliation{Institute of Theoretical Astrophysics, University of Oslo, Sem Sælands vei 13, 0371 Oslo, Norway}

\author{Nandan Roy}
\email{nandan.roy@mahidol.ac.th (Corresponding Author)} 
\affiliation{NAS, Centre for Theoretical Physics \& Natural Philosophy, Mahidol University,
Nakhonsawan Campus, Phayuha Khiri, Nakhonsawan 60130, Thailand}
% \date{\today}

\begin{abstract}
We study the late time cosmological dynamics of a two field dark energy model consisting of a canonical quintessence scalar field and a phantom scalar field in a spatially flat FLRW universe. The fields are minimally coupled to gravity and uncoupled at the level of the potential, with the quintessence sector governed by an exponential potential and the phantom sector by an inverse power law potential. By reformulating the background equations as a five dimensional autonomous dynamical system, we identify and analyze the fixed points and their stability properties, revealing stable late time attractors corresponding to phantom dominated accelerated expansion. We confront the model with observations through a Bayesian parameter estimation performed using the \textsc{Cobaya} framework, employing several combinations of recent cosmological data sets, including Pantheon+ supernovae, compressed cosmic microwave background distance priors, DESI DR2 baryon acoustic oscillation measurements, and DES Year-5 supernova data. The observational constraints favor a dynamical dark energy sector moderately and are consistent with deviations from a cosmological constant at the present epoch. The regions of parameter space preferred by the data are compatible with the stable accelerating solutions identified in the dynamical analysis, establishing a direct connection between phase space stability and observational viability. A notable feature of the model is that the effective dark energy equation of state undergoes phantom divide crossing in a gradual and asymptotic manner, rather than as a sharp transition.
\end{abstract}

\maketitle

\section{Introduction}

%The accelerated expansion of the universe, discovered through groundbreaking observations \cite{SupernovaSearchTeam:1998fmf, SupernovaCosmologyProject:1998vns, Meszaros:2002np, Planck:2014loa, ahn2012ninth}, is pivotal in modern cosmology. The \(\Lambda\)CDM model, which attributes this phenomenon to the cosmological constant \(\Lambda\), has long been regarded as the concordance or the standard model of cosmology due to its success in explaining a wide range of observations. However, precision-cosmology observations over the past decade have uncovered new tensions and discrepancies, prompting the search for alternative explanations.

Two major tensions in contemporary cosmology involve discrepancies in the measured values of the Hubble constant (\(H_0\)) and the amplitude of matter clustering (\(\sigma_8\)), as inferred from different cosmological observations. Early time probes, including the Cosmic Microwave Background (CMB) \cite{Planck2020}, Baryon Acoustic Oscillations (BAO) \cite{BAO2017}, Big Bang Nucleosynthesis (BBN) \cite{BBN2021}, DES \cite{DES2018}, estimate \(H_0\) to lie within \(67-68.5\) km/s/Mpc. These values are significantly lower, up to a \(5.3\sigma\) discrepancy, than late time measurements from SH0ES \cite{Sh0ES2019} and H0LiCOW \cite{Wong:2019kwg}, which find \(H_0 = 74.03 \pm 1.42\) km/s/Mpc \cite{Riess_2022}.  The parameter \(\sigma_8\), quantifies the root mean square of matter density fluctuations at a scale of \(8\ h^{-1}\) Mpc. A related parameter, \(S_8 = \sigma_8 \sqrt{\Omega_m/0.3}\), is often used. Low redshift probes, such as galaxy clustering, weak gravitational lensing, and galaxy cluster counts \cite{Simon:2022lde,Yuan:2022jqf,DES:2021vln,Busch:2022pcx,Wright:2025xka}, systematically yield lower \(S_8\) values compared to those derived from primary CMB anisotropies and the CMB lensing power spectrum, with a persistent \(2-3\sigma\) tension \cite{Planck:2015lwi}.

A recent challenge comes from the Dark Energy Spectroscopic Instrument (DESI), which, through full shape analysis of scale-dependent clustering of matter, suggests that dark energy may evolve with redshift rather than being a cosmological constant \cite{DESI:2024hhd, DESI:2024aqx, DESI:2024uvr, DESI:2025zgx, DESI:2025fii, Filbert:2023zhg, Ishak:2024jhs}. Furthermore, in \cite{Ishak:2024jhs}, combinations of these (first year) DESI clustering observations with cosmic microwave background (CMB) data from Planck, CMB lensing from Planck and ACT, Big Bang Nucleosynthesis (BBN) constraints on the baryon density, galaxy weak lensing and clustering from DESY3, and supernovae from DESY5 were shown to support general relativity while favoring a dynamical dark energy model with Chevallier–Polarski–Linder (CPL) equation of state parameters, $w_0 \neq -1$ and $w_a \neq 0$. However, the widely used CPL parameterization may be too simplistic to robustly probe the underlying microphysics of dark energy \cite{Cortes:2024lgw,Roy:2025cxk,Roy:2024kni,shlivko2024assessing}. Quintessence model with a thawing potential can, however, marginally accommodate these observations when mapped onto CPL parameters \cite{Wolf:2023uno, Shlivko:2024llw, Wolf:2024eph}. 

The inferred CPL parameters further suggest a dynamical dark energy equation of state (EoS) that crosses the phantom divide ($w = -1$), transitioning from $w < -1$ at higher redshifts to $w > -1$ today, with the crossing at $z_{\rm cross} \sim 0.4$–$0.5$ \cite{parametriccross, Lu:2025gki} (see also nonparametric reconstructions in \cite{adrianonparareconst}). However, the statistical significance typically exceeds $2\sigma$ but remains below the $5\sigma$ discovery threshold, rendering it inconclusive \cite{Keeley:2025rlg}. This apparent preference may partly reflect the uneven redshift coverage in the data or statistical fluctuations. Indeed, the inferred phantom behaviour could arise as an artefact of extrapolating the linear $w(a)$ form beyond the well constrained redshift range, while the underlying equation of state stays non-phantom throughout cosmic history \cite{Wolf:2023uno, Shlivko:2024llw, Wolf:2024eph, Patel:2024odo, Cortes:2024lgw}.

It is well established that these features cannot be accommodated within standard $\Lambda$CDM or single field quintessence models. Quintom models, consisting of two scalar fields, one canonical (quintessence-like) and one phantom like are therefore well motivated, as they allow the dark energy equation of state $w(z)$ to cross the phantom divide $w=-1$ \cite{Feng:2004ad, Guo:2004fq, Zhao:2005vj, Zhao2006,Goh:2025upc} (see \cite{Cai:2009zp} for a review). This has attracted significant recent interest, particularly in light of observational indications favouring dynamical dark energy and plausible phantom divide crossing \cite{cai2025quintom,li2025reconstructing, yang2025modified,tada2024quintessential, yang2024quintom, Gialamas:2025pwv, Goh:2025upc, Gomez-Valent:2025mfl}. In this work, we construct a quintom dark energy model where both fields are minimally coupled to gravity and interacting only through the background dynamics. An exponential potential governs the quintessence field, while a power-law potential governs the phantom field.  Their combined dynamics can allow the equation of state to cross the phantom divide. We have performed a detailed dynamical system analysis of this model by finding the fixed points and their corresponding stability. 
Finally, we compare the model with current cosmological data to constrain its parameters and assess its observational viability.

The rest of this paper is structured as follows. 
After briefly reviewing the no-go theorem for stable phantom crossing in single field dark energy models and highlighting the advantages of quintom scenarios, we introduce our model and derive the equations governing the background evolution in Sec.~\ref{sec:model}. In Sec.~\ref{The Dynamical System} we introduce the corresponding dynamical system. The fixed points of this system and their stability properties are examined in Sec.~\ref{Fixed points and stability}. In Sec.~\ref{Cosmological Evolution and Statefinder Analysis}, we investigate the evolution of various cosmological and cosmographic parameters, including a statefinder analysis. Sec.~\ref{Observational Data} describes the observational datasets and statistical techniques employed in this work. The resulting constraints from our data analysis are reported in Sec.~\ref{Constraints from Data Analysis}, and we close with a summary and outlook in Sec.~\ref{Conclusion}.

\section{Phantom Divide Crossing and Quintom Model}
\label{sec:model}

\subsection{Impossibility of Stable Phantom Crossing in single field Models}

A transition of the dark energy equation of state parameter through $w=-1$ is strongly constrained by fundamental consistency conditions. To illustrate this, consider a single scalar degree of freedom $\phi$ minimally coupled to gravity, with a general Lagrangian density
\begin{equation}
\mathcal{L} = F(X,\phi), \qquad
X \equiv \frac{1}{2}\nabla_\mu\phi\,\nabla^\mu\phi\  .
\end{equation}
The associated energy density and pressure are
\begin{equation}
\rho = 2X F_X - F, \qquad p = F \ ,
\end{equation}
yielding the effective equation of state
\begin{equation}
w = \frac{p}{\rho} = \frac{F}{2X F_X - F}\ .
\end{equation}

Reaching the phantom divide $w=-1$ requires the condition
$F_X=0$. However, the propagation of scalar fluctuations is governed by
the adiabatic sound speed,
\begin{equation}
c_s^2 \equiv \frac{\partial p/\partial X}{\partial \rho/\partial X}
= \frac{F_X}{F_X + 2X F_{XX}} \ .
\end{equation}
At the would be crossing point, the sound speed vanishes, while in the phantom regime $F_X<0$ implies $c_s^2<0$. The latter leads to exponential growth of short-wavelength perturbations, rendering the background solution dynamically unstable.

This obstruction is generic and does not depend on the detailed form of $F(X,\phi)$. Canonical quintessence, phantom scalar fields, and $k$-essence constructions all fail to realize a stable crossing within a single field description. This constitutes a robust no-go result for
phantom divide crossing with one propagating scalar degree of freedom.

\subsection{Role of Multiple Degrees of Freedom}

A minimal resolution is obtained by extending the dark energy sector to include more than one scalar mode. In the quintom framework considered here, the effective Lagrangian takes the form
\begin{equation}
\mathcal{L} = X_\phi - X_\sigma - V_{1} (\phi) - V_{2} (\sigma),
\end{equation}
where $\phi$ is canonical and $\sigma$ carries a negative kinetic term.  Each field separately obeys

\begin{equation}
w_\phi > -1, \qquad w_\sigma < -1 ,
\end{equation}
while the total dark energy equation of state,
\begin{equation}
w_{\rm DE} = \frac{p_\phi + p_\sigma}{\rho_\phi + \rho_\sigma}\ ,
\end{equation}

is free to evolve smoothly across $w=-1$. The crossing is controlled by the relative contribution of the kinetic sectors and does not require either field to violate its individual stability bound.

It is worth emphasizing that the two-field quintom model is not the only viable approach to achieving phantom divide crossing. A combination of standard and negative quintessence was also shown to feature phantom divide crossing \cite{Gomez-Valent:2025mfl}. Alternative realizations include single field models with non-canonical (e.g., k-essence) or higher-derivative kinetic terms in the Horndeski class \cite{Armendariz2001, Sur:2008tc, Saitou:2012xw, Matsumoto:2017qil}, hessence models based on a complex scalar field with internal degrees of freedom \cite{Wei:2005nw}, braneworld scenarios \cite{Wu:2007fx}, modified gravity theories like $f(R)$ gravity \cite{Bamba:2010zxj, Nojiri:2025low}, interacting dark energy models \cite{Roy:2018eug,Kritpetch:2024rgi,Sahoo:2025cvz}.

\subsection{The Model}\label{The Model}

We consider a universe that is spatially flat and has both a canonical and a phantom scalar field representing the dark energy sector. Then the Friedmann and acceleration equations are respectively given by

\begin{equation}\label{rsh01}
    H^2 = \frac{\kappa^2}{3} \left( \rho_m + \frac{1}{2} \dot{\phi}^2 - \frac{1}{2} \dot{\sigma}^2 + V_{1}(\phi)+V_{2}(\sigma) \right) \ ,
\end{equation}

\begin{equation}\label{rsh02}
\dot{H} = -\frac{\kappa^2}{2} \left( \rho_m + \dot{\phi}^2 - \dot{\sigma}^2 \right) \ .
\end{equation}

%  \begin{equation}\label{rsh02}
% \dot{H}=-\frac{\kappa^{2}}{2}[\left( p_{m}+\rho_{m}\right)+\left( p_{r}+\rho_{r}\right)+\left(p_{\phi}+\rho_{\phi}\right)+\left( p_{\Lambda}+\rho_{\Lambda}\right)].
% \end{equation}

Here, $\kappa^{2}=8\pi G$, with $H=\dot{a}/a$ denoting the Hubble function and $a(t)$ representing the scale factor. The quantities related to matter, quintessence, and the phantom will be denoted through respective subscripts $m$, $\phi$, and $\sigma$, and the overdots indicate differentiation with respect to the cosmic time.  

We have standard continuity equations $\dot{\rho_{i}}=-3H(p_{i}+\rho_{i})$ for each species $i \in \{m,\phi,\sigma\}$. The Klein-Gordon equation for the scalar fields can be expressed as follows:

\begin{equation}\label{rsh03}
\ddot{\phi}+3H\dot{\phi}+\frac{dV_{1}(\phi)}{d\phi}=0 \ , 
\end{equation}

\begin{equation}\label{rsh03}
\ddot{\sigma}+3H\dot{\sigma}-\frac{dV_{2}(\sigma)}{d\sigma}=0 \ , 
\end{equation}

In this work, we adopt the following scalar field potentials:

\begin{equation}
\begin{split}
    V_{1}(\phi) &= V_{\phi 0} e^{-\lambda_{\phi} \phi}\ ,\\
    V_{2}(\sigma) &= M^{4+m}(\sigma)^{-m} \quad \text{(with $m> 0$)}\ .
\end{split}
\end{equation}

The steep exponential potential for the quintessence field does not support slow roll evolution and therefore keeps the field subdominant at early times and becomes dynamical at late times. Such potentials constitute a standard realization of thawing dark energy models, in which the equation of state $w$ remains close to $-1$ during the matter dominated era and subsequently evolves toward less negative values at late times \cite{Andriot:2024sif, Andriot:2025los, DESI:2025fii}. The phantom field is governed by an inverse power-law potential, which becomes naturally flat for large field values and admits a late time attractor solution. Power-law potentials for phantom fields are particularly well motivated, as they can give rise to tracker behavior, thereby reducing sensitivity to initial conditions \cite{Zhao2006,LinaresCedeno:2021aqk}. Unlike Ref.~\cite{Saridakis:2009pj}, where direct power-law potentials were considered, here we focus on inverse power-law forms. Such potentials also arise in supersymmetric constructions involving phantom degrees of freedom \cite{SUSYphatom1, SUSYphantom2}. 

In the numerical analysis presented in this work, we fix the exponent of the phantom potential to $m=4$.  Our goal here is not to perform a full exploration of the parameter space of the phantom potential but rather to illustrate the cosmological dynamics of the quintom scenario presented in this work and its observational viability.

\subsection{Linear Scalar Perturbations}

We now briefly comment on perturbative behavior of the model. Scalar perturbations around a spatially flat FLRW background can be described in the conformal Newtonian gauge,
\begin{equation}
ds^2 = a^2(\tau)\big[(1+2\Phi)d\tau^2 - (1-2\Phi)\delta_{ij}dx^i dx^j\big],
\end{equation}
where the equality of the metric potentials follows from the absence of anisotropic stress.

For each scalar component $i\in\{\phi,\sigma\}$, the linearized energy conservation equations take the form \cite{Ma:1995ey,Goh:2025upc}
\begin{align}
\dot{\delta}_i &= -(1+w_i)(\theta_i - 3\dot{\Phi})
- 3\mathcal{H}(c_{s,i}^2 - w_i)\delta_i , \\
\dot{\theta}_i &= -\mathcal{H}(1-3w_i)\theta_i
- \frac{\dot{w}_i}{1+w_i}\theta_i
+ \frac{c_{s,i}^2}{1+w_i}k^2\delta_i + k^2\Phi .
\end{align}

Unlike single field models, no divergence arises at $w_{\rm DE}=-1$, since neither $\phi$ nor $\sigma$ individually crosses the phantom boundary. The effective dark energy perturbations are obtained through energy-weighted combinations \cite{Zhao:2005vj},
\begin{subequations}
\begin{align}
\delta_{\rm DE} &=
\frac{\rho_\phi\delta_\phi + \rho_\sigma\delta_\sigma}
{\rho_\phi + \rho_\sigma},  \\
\qquad
\theta_{\rm DE} &=
\frac{(\rho_\phi+p_\phi)\theta_\phi + (\rho_\sigma+p_\sigma)\theta_\sigma}
{\rho_\phi+p_\phi + \rho_\sigma+p_\sigma}.
\end{align}
\end{subequations}

It has been shown in \cite{Goh:2025upc, Zhao:2005vj}, in a set up like ours, both fields propagate with unit sound speed, $c_{s,\phi}^2 = c_{s,\sigma}^2 = 1$, ensuring the absence of gradient instabilities. Consequently, the effective perturbations remain finite through the phantom divide, allowing a consistent realization of phantom crossing at the perturbation level.

\section{The Dynamical System }\label{The Dynamical System}

To understand the phase space behaviour of the system, one needs to introduce a new set of dimensionless variables to write it as an autonomous system. Here we consider the following set of dimensionless transformations:

\begin{align}\label{rsh06}
 x_{\phi} ^{2} & = \frac{\kappa^{2}\dot{\phi}^{2}}{6H^{2}}\ , \quad y_{\phi}^{2}= \frac{\kappa^{2}V_1 (\phi)}{3H^{2}},\ 
\end{align}

\begin{align}\label{rsh06}
 x_{\sigma} ^{2} & = \frac{\kappa^{2}\dot{\sigma}^{2}}{6H^{2}}\ , \quad y_{\sigma}^{2}= \frac{\kappa^{2}V_2 (\sigma)}{3H^{2}}\ ,
\end{align}

\begin{align}
    \lambda_{\phi} = -\frac{1}{\kappa V_{1}(\phi)} \frac{dV_{1}(\phi)}{d\phi}\ , \quad \lambda_{\sigma} = -\frac{1}{\kappa V_{2}(\sigma)} \frac{dV_{2}(\sigma)}{d\sigma} \ .
\end{align}

With the help of the above transformations, the background field equations can be reduced to a set of autonomous equations for the choice of the potentials mentioned before:

\begin{subequations}\label{eq:autonomous1}
\begin{align}
    x_\phi' &= -3x_\phi + \sqrt{\frac{3}{2}} \lambda_\phi y_\phi^2 + \frac{3}{2} x_\phi~ \mathcal{U}\ ,\label{eq:x1} \\
    x_\sigma' &= -3x_\sigma - \sqrt{\frac{3}{2}} \lambda_\sigma y_\sigma^2 + \frac{3}{2} x_\sigma ~\mathcal{U} \ ,\label{eq:x2}  \\
    y_\phi^\prime &= -\sqrt{\frac{3}{2}}  \lambda_\phi x_\phi y_\phi + \frac{3}{2} y_\phi~ \mathcal{U}\ ,\label{eq:y1}\\
    y_\sigma^\prime &= -\sqrt{\frac{3}{2}}  \lambda_\sigma x_\sigma y_\sigma + \frac{3}{2} y_\sigma~ \mathcal{U}\ ,\label{eq:y2} \\
    \lambda_\sigma' &= -\sqrt{6} \lambda_\sigma^2 x_\sigma (\frac{1}{m}) ,\\
    \text{where, } \ \mathcal{U} &= \left( 1 + x_\phi^2 - x_\sigma^2 - y_\phi^2 - y_\sigma^2 \right)\ .
\end{align}
\end{subequations}

The prime is the differentiation with respect to  $N= \ln a$. Different cosmological variables can be expressed in terms of the dynamical system variables as follows, 

\begin{equation}
    \Omega_\phi = x_\phi^2 + y_\phi^2, \qquad \Omega_{\sigma} = -x_\sigma^2+y_\sigma^2\ ,
\end{equation}

\begin{equation}
    \Omega_{\rm DE} = x_\phi^2-x_\sigma^2+y_\phi^2+y_\sigma^2\ ,
\end{equation}

\begin{equation}
     w_{\phi} =\frac{x_\phi^2-y_\phi^2}{x_\phi^2+y_\phi^2}\ , \quad w_{\sigma} =\frac{-x_\sigma^2-y_\sigma^2}{-x_\sigma^2+y_\sigma^2}\ ,
\end{equation}

\begin{equation}\label{rsh07}
 w_{\rm DE} \equiv \frac{p_{\rm DE}}{\rho_{\rm DE}}=\frac{x_\phi^2-x_\sigma^2-y_\phi^2-y_\sigma^2}{x_\phi^2-x_\sigma^2+y_\phi^2+y_\sigma^2}\ , 
\end{equation}

\begin{equation}\label{rsh07}
q = -1+\frac{3}{2}{(1+x_\phi^2-x_\sigma^2-y_\phi^2-y_\sigma^2)}\ .
\end{equation}

Here, $\Omega_{\phi}$, $\Omega_{\sigma}$ and $\Omega_{\rm DE}$ represents the quintessence, phantom and total dark energy density parameter respectively with $w_{\phi}$, $w_{\sigma}$ and $w_{\rm DE}$ as the corresponding equation of state while $q$ is the universe's deceleration parameter. In the next section, we analyze the fixed points and their stability.

\begin{table*}[ht]
\centering
\begin{tabular}{|c|c|c|c|c|c|c|}
\hline
$\text{Fixed Point}$ & $x_\phi$ & $x_\sigma$ & $y_\phi$ & $y_\sigma$ & $\lambda_\sigma$ & Eigenvalues\\ \hline
$p_1$ & $0$ & $0$ & $0$ & $0$ & -- & $\left\{-\frac{3}{2},-\frac{3}{2},\frac{3}{2},\frac{3}{2},0\right\}$ \\ \hline

$p_2$ & $\pm 1$ & $0$ & $0$ & $0$ & -- & $\left\{3,3,0,0,\frac{1}{2} \left(\sqrt{6} \lambda _{\phi } \mp 6\right)\right\}$ \\ \hline

$p_3$ & $\frac{\sqrt{3/2}}{\lambda_\phi}$ & $0$ & $\pm\frac{\sqrt{3/2}}{\lambda_\phi}$ & $0$ & -- & $\left\{0,-\frac{3}{2},\frac{3}{2},\frac{3 \left(-\lambda _{\phi }^2-\sqrt{24 \lambda _{\phi }^2-7 \lambda _{\phi }^4}\right)}{4 \lambda _{\phi }^2},\frac{3 \left(\sqrt{24 \lambda _{\phi }^2-7 \lambda _{\phi }^4}-\lambda _{\phi }^2\right)}{4 \lambda _{\phi }^2}\right\}$\\ \hline

$p_4$ & $\frac{\lambda_\phi}{\sqrt{6}}$ & $0$ & $\pm\frac{\sqrt{6 - \lambda_\phi^2}}{\sqrt{6}}$ & $0$ & -- & $\left\{0,\frac{\lambda _{\phi }^2}{2},\frac{1}{2} \left(\lambda _{\phi }^2-6\right),\frac{1}{2} \left(\lambda _{\phi }^2-6\right),\lambda _{\phi }^2-3\right\}$ \\ \hline

$p_5$ & $0$ & $0$ & $0$ & $\pm1$ & $0$ & ${-3,-3,-3,0,0}$\\ \hline

\end{tabular}
\caption{ The table of fixed points for the system of equations Eq.~(\ref{eq:autonomous1}) with corresponding eigenvalues.}
\label{tab:fixed_points}
\end{table*}

\begin{table*}[ht]
\centering
\small
\begin{tabular}{|c|c|c|c|c|c|c|}
\hline
Fixed point & $\Omega_\phi$ & $\Omega_\sigma$ & $\Omega_m$  & $w_{DE}$ & $q$ & Existence Condition \\
\hline
$p_1$ & 0 & 0 & 1 & -1 : ($x_i, y_i \rightarrow 0$) & $\frac{1}{2}$ & always \\
\hline
$p_2$ & 1 & 0 & 0 & 1 & 2 & always\\
\hline
% $P_4$ & 0 & $\sqrt{1-x_\phi^2}$ & $1-\sqrt{1-x_\phi^2}$ & 1 & 2 \\ 
% \hline 
$p_3$  & $\frac{3}{\lambda_{\phi}^{2}}$ & 0 & $1-\frac{3}{\lambda_{\phi}^{2}}$  & 0 & $\frac{1}{2}$ & $\lambda_{\phi}\geq \sqrt{3} \quad (0\leq \Omega_\phi \leq 1)$ \\ 
\hline
$p_4$  &  1 & 0 &  0  & $\frac{\lambda_\phi^2 - 3}{3}$ & $-1+\frac{\lambda_\phi^2}{2}$ &$\lambda_{\phi}\leq \sqrt{6}$ \\ 
\hline
$p_5$  & 0 & 1 & 0 & -1 & -1 & always\\

% $p_6$  & $\frac{\sqrt{3/2}}{\lambda_\phi}$ & 0 & $1-\frac{\sqrt{3/2}}{\lambda_\phi}$ & 0 & $\frac{1}{2}$ \\  
% \hline
% $p_6$  & \textcolor{red}{$\frac{\sqrt{6}}{\lambda_\phi}$}   \textcolor{blue}{$\frac{\lambda_{\phi}^{2}}{6}$} & \textcolor{red}{$\frac{\sqrt{6-\lambda_\phi^2}}{\lambda_\phi}$} \textcolor{blue}{$-1+\frac{\lambda_{\phi}^{2}}{6}$} & \textcolor{red}{$1-(\frac{\sqrt{6}+\sqrt{6-\lambda_\phi^2}}{\lambda_\phi})$} \textcolor{blue}{$2(-1+\frac{\lambda_{\phi}^{2}}{6})$} & 1 & 2 \\ 
\hline
% $p_{8}$  & $\frac{\lambda_\phi}{\sqrt{6}}$ & 0 & $1-\frac{\lambda_\phi}{\sqrt{6}}$ & $\frac{\lambda_\phi^2-3}{3}$ & $-1+\frac{\lambda_\phi^2}{2}$ \\ 
\end{tabular}
\caption{List of fixed Points with their cosmological behaviours and corresponding existance condition.}\label{tab:fixedpoint2}
\end{table*}

\section{Fixed points and stability}\label{Fixed points and stability}

The fixed points of the system are determined by solving Eq.~(\ref{eq:autonomous1}) with all derivatives with respect to $N$ set to zero. The resulting fixed points and their associated eigenvalues are summarized in Table~\ref{tab:fixed_points}. The corresponding physical characteristics and existance condition of these fixed points are provided in Table~\ref{tab:fixedpoint2}.

The stability of a fixed point in a nonlinear system is determined by the eigenvalues of the Jacobian matrix evaluated at that point. When all the eigenvalues have non-zero real parts (i.e., the fixed point is hyperbolic), linear stability analysis can be directly applied. However, in the case of non-hyperbolic fixed points, where one or more eigenvalues have zero real parts, stability must be analysed using analytical tools such as the centre manifold theorem or suitable numerical techniques. For a review on the dynamical systems in cosmology, see \cite{Bahamonde:2017ize}. 

The fixed points $p_1$ correspond to matter dominated solutions with $\Omega_m=1$ and the corresponding value of the deceleration parameter is $q=\frac{1}{2}$. Although these points are non-hyperbolic due to the presence of a zero eigenvalue, the coexistence of both positive and negative eigenvalues indicates that they are saddle points in phase space, as expected for a transient matter dominated epoch. 

The fixed point $p_{2}$ represents a universe dominated by the kinetic energy of the quintessence field, with $x_\phi=\pm1$. In this regime, the scalar field behaves as a stiff fluid with equation of state $w=1$, leading to strong deceleration ($q=2$). This point is unstable and corresponds to an early time kinetic dominated phase for the quintessence scalar field.

The fixed point $p_{3}$ corresponds to a scaling solution in which the quintessence field and matter coexist. In this case, the effective equation of state of the quintessence field coincides with that of matter, $w_\phi=0$, making the scalar field dynamically indistinguishable from dust. This point is decelerating and unstable, and therefore can only represent a transient phase of the cosmic evolution.

The fixed point $p_{4}$ describes a scenario completely dominated by the quintessence field. This fixed point exist for $\lambda_{\phi}\leq \sqrt{6}$. Although this point can yield accelerated expansion for $\lambda_\phi^2<2$, it possesses both positive and negative eigenvalues and is therefore a saddle point. Consequently, despite allowing for acceleration of the universe, it cannot act as a late time attractor. 

The fixed point $p_{5}$ is characterized by complete domination of the potential energy of the phantom field, with $y_\sigma=\pm1$. This solution corresponds to a de Sitter like accelerated phase with $w=-1$ and $q=-1$. The eigenvalue spectrum consists of three negative eigenvalues and two vanishing eigenvalues, rendering the point non-hyperbolic. Consequently, linear stability analysis alone is insufficient to establish its asymptotic behaviour.

\begin{figure*}
    \centering
    \includegraphics[width=1\linewidth]{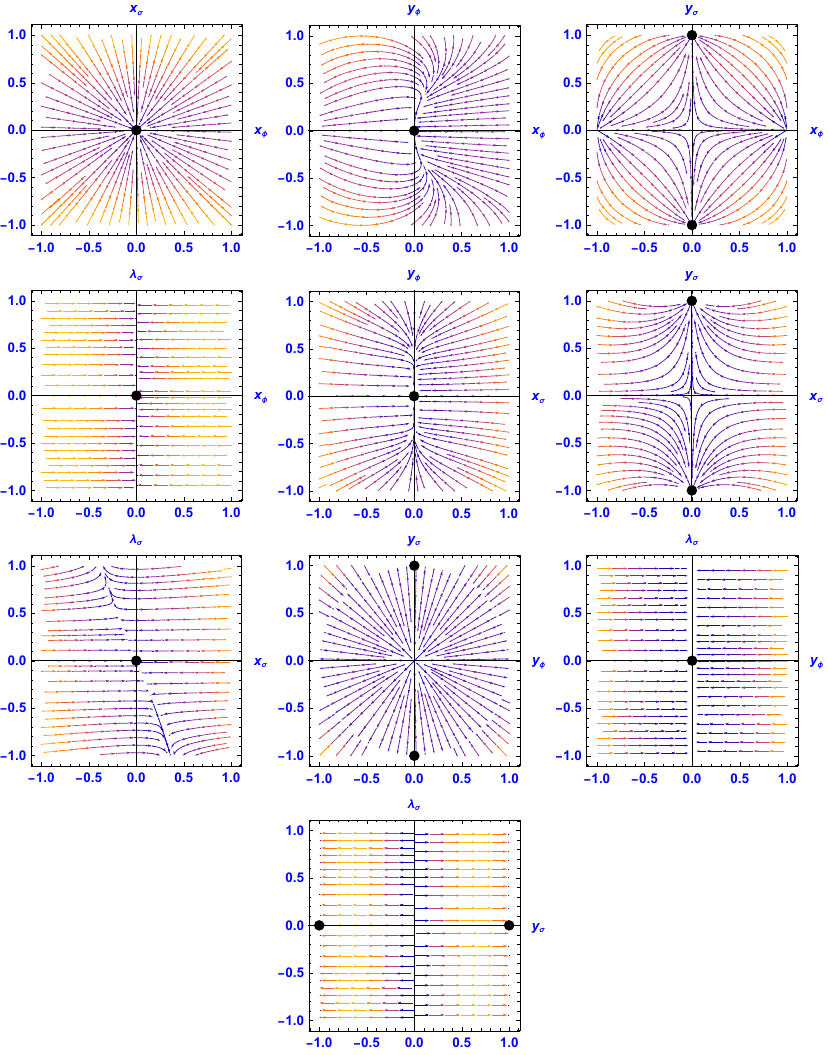}
    \caption{The two-dimensional projected phase spaces corresponding to the fixed point $p_{5}$. In each projected phase space, the projection of the fixed point is marked by a black dot. All plots are generated for $\lambda_{\phi}=1.5$ and $m=4$.}
    \label{fig:phaseplot}
\end{figure*}

To assess the stability of $p_{5}$, the phase-space trajectories are investigated numerically by constructing two-dimensional projected phase spaces for all combinations of the dynamical variables. For the numerical computation, we consider $\lambda_{\phi}=1.5$, corresponding to the posterior values reported in Table~\ref{tab:bestfit} and $m=4$. In addition, $y_{\sigma}=1$ is fixed for the phase planes that do not explicitly contain $y_{\sigma}$. The resulting phase portraits, displayed in Fig.~\ref{fig:phaseplot}, show that trajectories in the physically relevant region of the phase space are attracted toward the critical point $p_{5}$. This behavior indicates that $p_{5}$ can act as a stable late-time attractor in the full phase space at the level of background cosmological dynamics.

% The fixed point \textcolor{red}{\sout{$p_7$}} \textcolor{blue}{$p_{6}$} represents a special case of the scaling solution $p_4$, corresponding to a restricted subspace of the full phase space. Although $p_7$ can be formally treated as a separate critical point, it shares the same physical interpretation, cosmological properties, and stability behaviour as $p_4$.

% The fixed point \textcolor{blue}{$p_6$} corresponds to complete kinetic domination by both the quintessence and phantom fields. \textcolor{blue}{This fixed point exist for $\lambda_{\phi}\leq \sqrt{6}$}. Despite the presence of zero eigenvalues, the positive eigenvalues render this point unstable. Since the effective equation of state is again $w=1$, this solution always leads to decelerated expansion and cannot describe either late time acceleration or a viable matter dominated epoch.

% The fixed point $p_{9}$ corresponds to a special case of the nondegenerate fixed point $p_5$, obtained by restricting the system to $\lambda_\sigma=0$. Its phase–space coordinates, cosmological parameters, and eigenvalue structure coincide with those of $p_5$. Consequently, $p_{9}$ represents the same physical configuration, namely a quintessence-dominated universe that can exhibit accelerated expansion for $\lambda_\phi^2<2$, but which is dynamically unstable due to the presence of positive eigenvalues. As in the case of $p_5$, this fixed point can at most describe a transient accelerating phase and cannot act as a late time attractor. 

The dynamical systems analysis performed here reveals that, among all fixed points of the model, only the phantom potential–dominated solution $p_5$ can act as a late time attractor capable of driving cosmic acceleration. All other fixed points correspond to either decelerated or unstable phases and therefore represent transient stages in the cosmic evolution.

\section{Cosmological Evolution and Statefinder Analysis}\label{Cosmological Evolution and Statefinder Analysis}

\begin{figure}
  \centering
  \begin{subfigure}{\linewidth}
    \includegraphics[width=\linewidth]{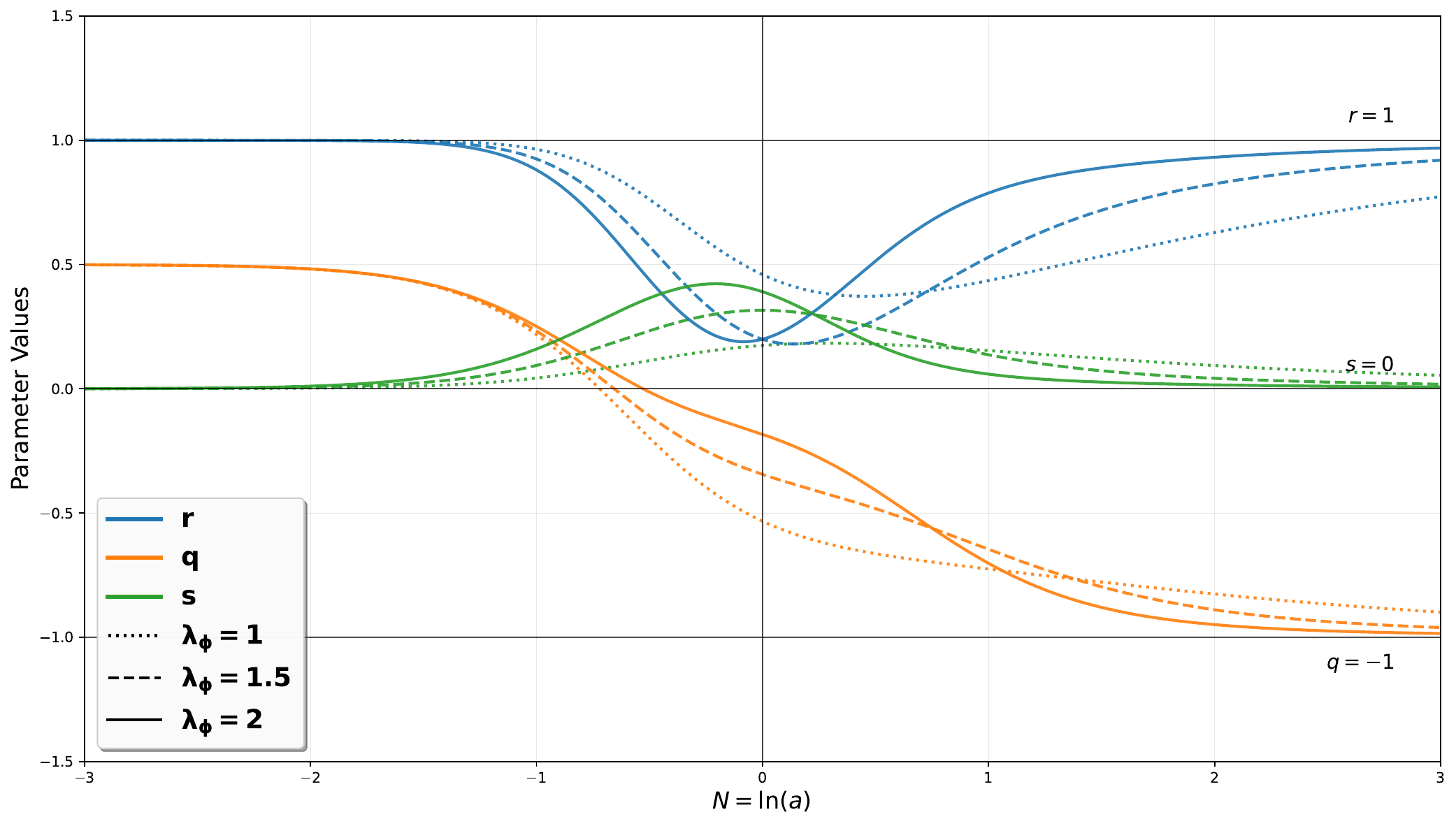}
    \caption{The plot of the deceleration parameter ($q$) in orange, state finder parameter $r$ in blue and the $s$ in green.}\label{fig:rqs}
  \end{subfigure}
  \vspace{6pt}
  \begin{subfigure}{\linewidth}
    \includegraphics[width=\linewidth]{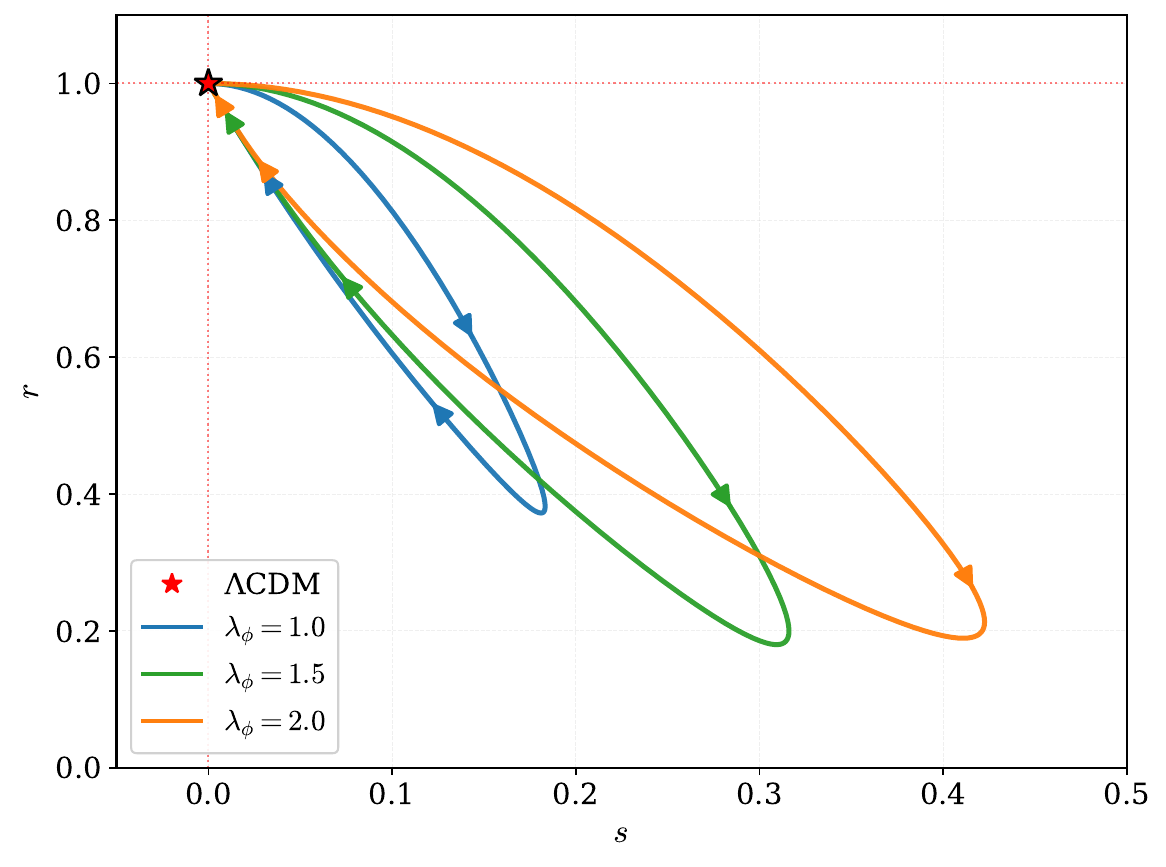}
    \caption{Parametric plot of the solutions in the statefinder $r - s$ plane.}\label{fig:statefindrqs}
  \end{subfigure}
  \vspace{6pt}
  \begin{subfigure}{\linewidth}
    \includegraphics[width=\linewidth]{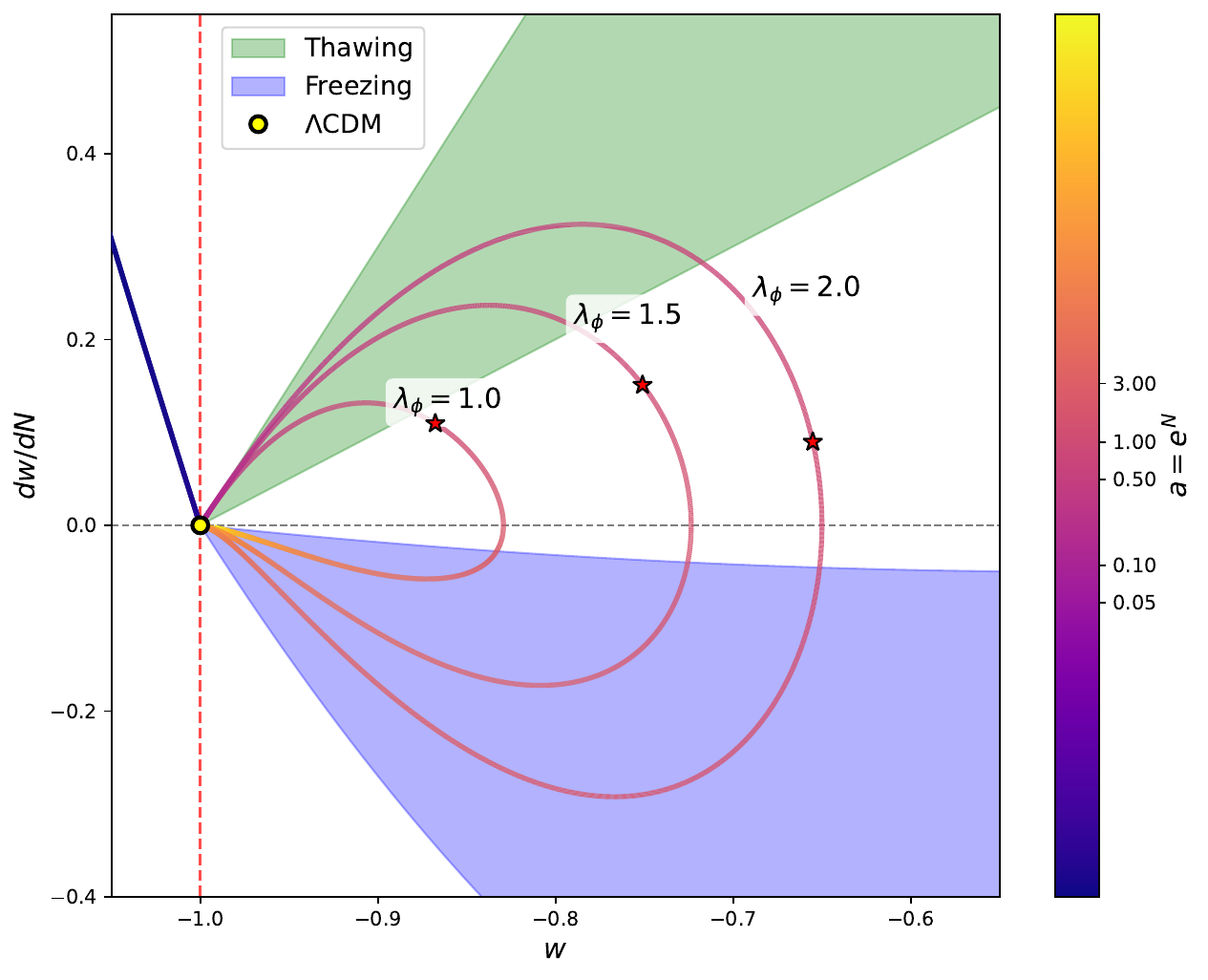}
    \caption{The $w'_{\rm DE} - w_{\rm DE}$ phase space with the present epoch ($a = 1$) represented by the star.}\label{fig:thaw_freeze}
  \end{subfigure}
  \caption{Plot of different cosmographic and state finder parameters, $q,r,s$ (top panel), $r-s$ phane plot (middle panel) and $\frac{dw}{dN}$ vs $w$ (bottom panel). The dotted, dashed, solid curves represent $\lambda_\phi = 1, 1.5, 2$ respectively.}
  \label{fig:statefinder}
\end{figure}

In this section, we discuss the cosmological evolution of the quintom model and analyze its dynamical properties using the statefinder diagnostic. The discussion is based on Figs.~\ref{fig:Density_eos}--\ref{fig:statefinder}, obtained by numerically integrating the full background equations. Though in the dynamical system analysis in the previous section, we have not included radiation since we are interested in the late time dynamics, but to solve the system of equations numerically, we have considered radiation in the system and have set the initial condition in the radiation dominated era, such that we can obtain all the cosmological eras properly.

To reproduce a dark energy equation of state (EoS) crossing the phantom divide in the recent past, with the EoS evolving from the phantom regime to the quintessence regime, we require the effective EoS to start in the phantom region at early times. This can be achieved by choosing initial conditions such that $x_{\phi_i} < x_{\sigma_i}$, while keeping all other dynamical system variables of the same order of magnitude. In addition, the condition $x_{\sigma_i} < y_{\sigma_i}$ must be satisfied to ensure that the phantom field energy density remains physical.

Subject to these requirements, we adopt the following initial conditions at $N = -8.5$: $x_{\phi_i}= 0.2\times 10^{-7}, x_{\sigma_i} = 0.1\times 10^{-5}, y_{\phi_i} = 0.35\times 10^{-5}, y_{\sigma_i} = 0.2\times 10^{-5}, \lambda_{\sigma_i}=0.05, \Omega_r=0.999$.  We have verified that small variations of these
initial conditions do not qualitatively change the cosmological evolution.
This behaviour is expected from the dynamical systems perspective, since
the radiation and matter dominated solutions correspond to saddle points
of the system that act as quasi-attractors, guiding the trajectories in phase space for a prolonged period before the system eventually evolves
toward the late time phantom dominated attractor. The values of $\lambda_\phi$ are chosen in accordance with the constraints obtained from the MCMC analysis presented in the next section. For all the plots and the MCMC analysis presented here we have considered $m=4$.

\subsection{Background Evolution and Phantom Crossing}
\begin{figure}
  \centering
  \begin{subfigure}{\linewidth}
    \includegraphics[width=\linewidth]{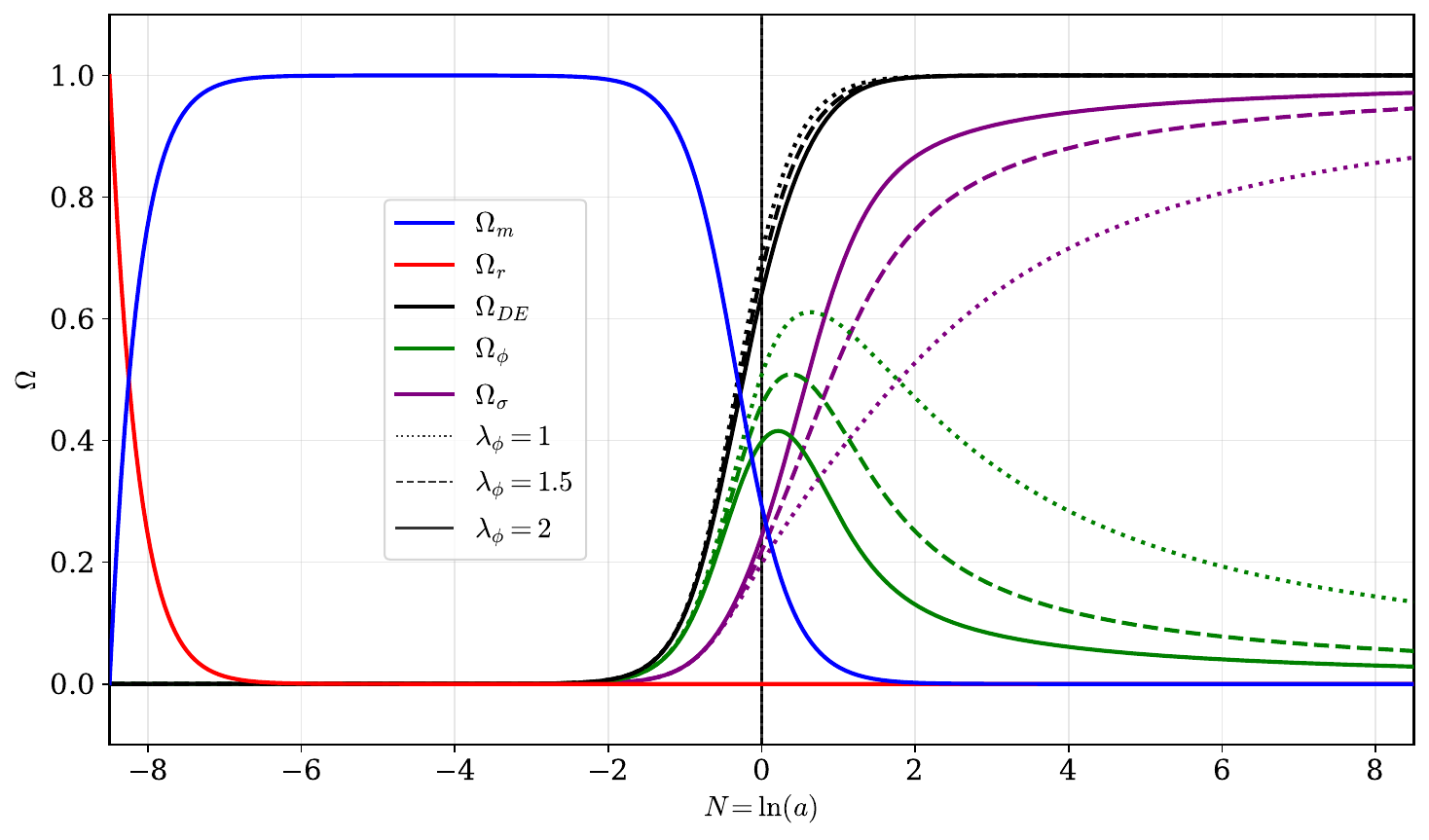}
    \caption{Evolution of density fractions $\Omega_m, \Omega_r, \Omega_\phi, \Omega_\sigma$ and $\Omega_{DE}$}\label{fig:density.}
  \end{subfigure}
  \vspace{6pt}
  \begin{subfigure}{\linewidth}
    \includegraphics[width=\linewidth]{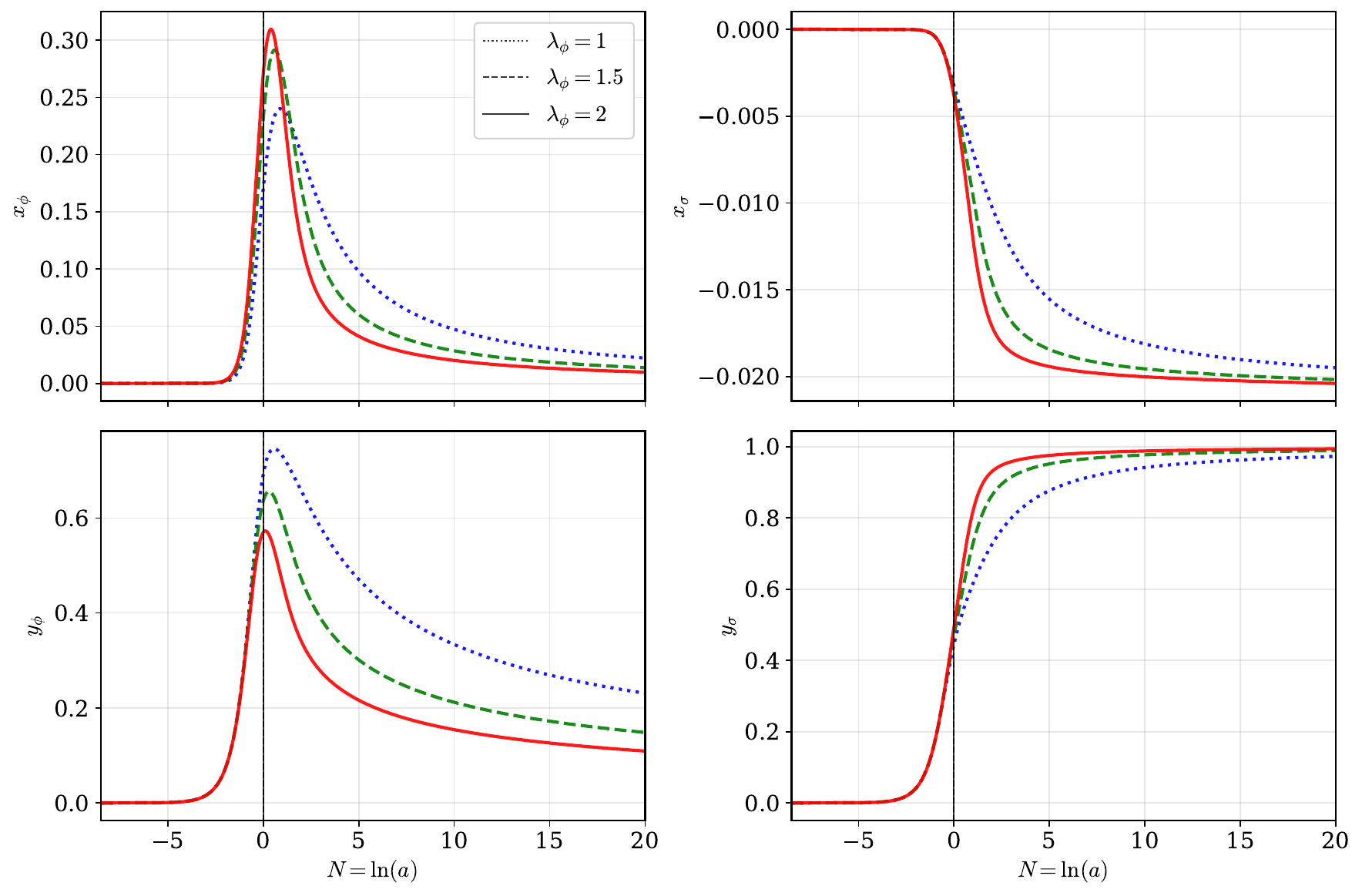}
    \caption{Plot of the evolution of the dimensionless variables $x_\phi$ (top left), $y_\phi$ (bottom left), $x_\sigma$ (top right), $y_\sigma$ (bottom right).}\label{fig:dynamicvar}
  \end{subfigure}
  \vspace{6pt}
  \begin{subfigure}{\linewidth}
    \includegraphics[width=\linewidth]{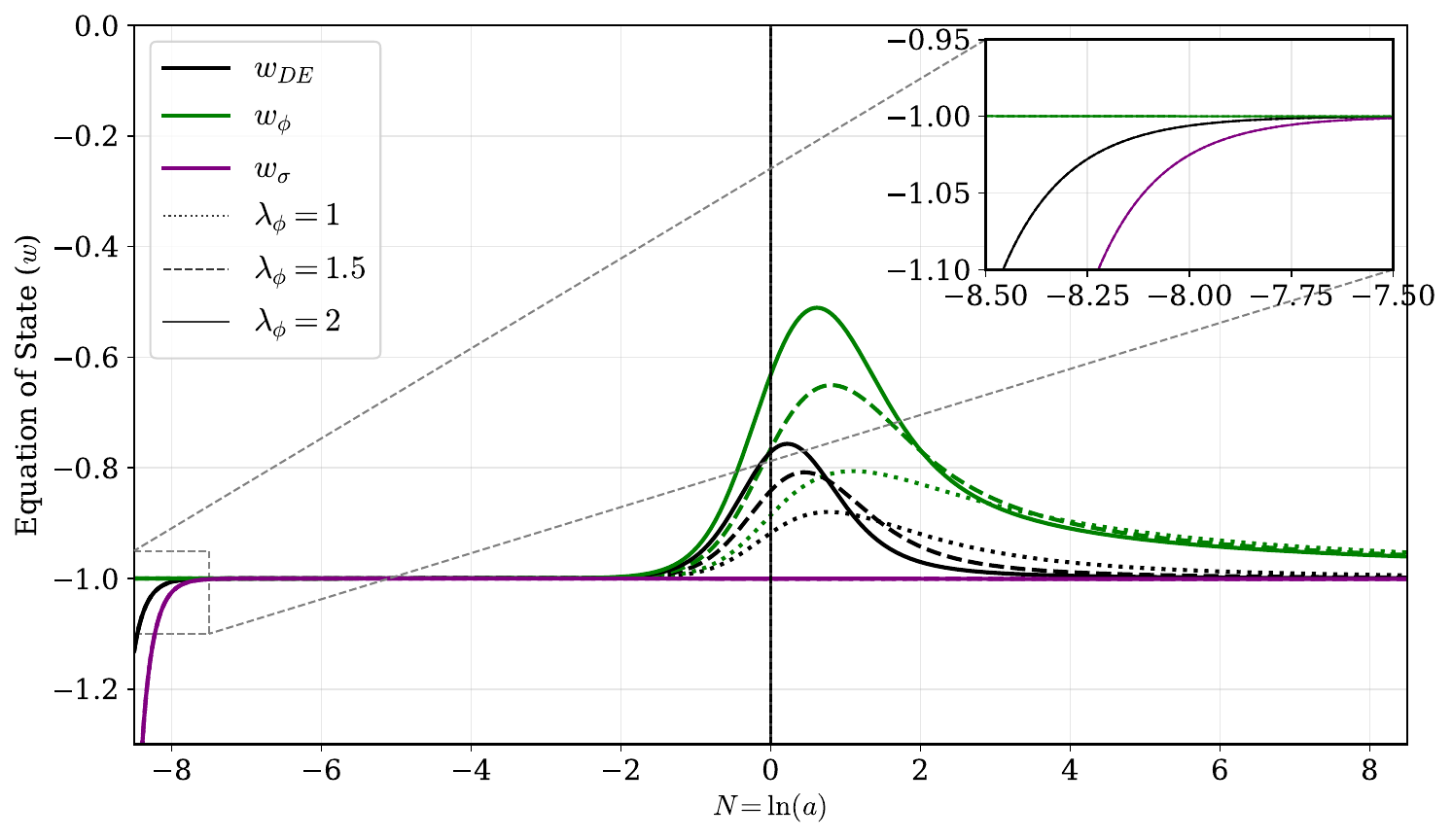}
    \caption{Plot of the equation of state of the scalar fields, quintessence ($w_\phi$) in green, phantom ($w_\sigma$) in purple and the effective dark energy EoS ($w_{DE}$ in black). }\label{fig:eos}
  \end{subfigure}
  \caption{Plot of different cosmological parameters, density fractions (top panel), dimensionless dynamical system variable (middle panel) and equation of states (bottom panel). The dotted, dashed, solid curves represent $\lambda_\phi = 1, 1.5, 2$ respectively.}
  \label{fig:Density_eos}
\end{figure}

Fig.~\ref{fig:density.} shows the evolution of the density parameters $\Omega_m$, $\Omega_r$, and $\Omega_{\rm DE}$, together with $\Omega_\phi$ and $\Omega_\sigma$, as functions of the e-folding number $N=\ln a$ for three representative values of the quintessence slope parameter $\lambda_\phi = \{1, 1.5, 2\}$. The universe exhibits the standard sequence of cosmological epochs, beginning with radiation domination, followed by a prolonged matter dominated era, and finally transitioning to dark-energy domination at late times. It can also be seen that the present day values of the density parameters inferred from these plots are consistent with the posterior constraints obtained
in the next section. The evolution of the dimensionless variables $x_\phi$, $x_\sigma$, $y_\phi$, and $y_\sigma$ is shown in Fig.~\ref{fig:dynamicvar}.

In Fig.~\ref{fig:eos}, we plot the evolution of the equations of state of the scalar fields, $(w_\phi, w_\sigma)$, together with the total dark energy equation of state, $w_{\rm DE}$, for the same values of $\lambda_\phi$ used in the density parameter plot. For the choice of initial conditions specified earlier, $w_{\rm DE}$ initially lies in the phantom regime, rapidly approaches $w_{\rm DE} \simeq -1$, and remains nearly constant throughout the matter dominated era. With the onset of dark-energy domination, $w_{\rm DE}$ becomes dynamical and evolves into the quintessence regime, crossing the phantom divide. In future, it asymptotically returns to $w_{\rm DE} \simeq -1$, effectively mimicking a cosmological constant.

From Fig.~\ref{fig:dynamicvar}, we observe that the quintessence field is frozen in the early universe, becomes dynamical at late times during the onset of dark-energy domination, and freezes again in the future. Both the kinetic ($x_\phi$) and potential ($y_\phi$) contributions of the
quintessence field asymptotically vanish. In contrast, the phantom field becomes dynamical at late times, and in the asymptotic future the universe is dominated by the phantom sector with $y_\sigma \simeq 1$. This behavior confirms our dynamical system analysis in the previous section, which predicts a single late time attractor dominated by the phantom field.

It is worth noting, however, that although the total dark energy equation of state $w_{\rm DE}$ starts in the phantom regime in the very early universe, the visible phantom divide crossing occurs at $z \simeq 1$--$4$. According to the current model, the phantom crossing may therefore occur earlier than suggested by the DESI DR2 observations. Moreover, the crossing takes place gradually and asymptotically, rather than as a sharp transition\cite{Ma:1995ey,Roy:2024kni,Roy:2023vxk}. This behaviour  can be justified from the dynamical system perspective. In general during the matter dominated era the dynamical variables typically evolve towards the neighbourhood of the matter dominated saddle points of the system, where $x_i \to 0$ and $y_i \to 0$. For the class of initial
conditions considered in our numerical analysis this implies that the effective dark energy equation of state remains close to $w_{\rm DE}\simeq -1$ throughout the matter dominated epoch.

However, this behaviour is not strictly universal and can depend on the choice of parameters and initial conditions. In particular, for proper choice of the quintessence slope parameter $\lambda_\phi$ the system may temporarily approach the scaling solution $p_3$, in which the quintessence field tracks the matter component and the effective dark energy equation of state can deviate from $-1$ and move closer to $w_{\rm DE}\sim 0$.

The evolution of the Hubble parameter $H(N)$ is compared with observational data from cosmic chronometers (see the appendix of Ref.~\cite{Sahoo:2024dgb}) in Fig.~\ref{fig:Hubble}, using the same initial conditions as in the previous plots. From the figure, it is evident that the quintom model provides a good fit to the observed Hubble data at late times.

\begin{figure}[!hbt]
            \centering
            \includegraphics[width=\columnwidth]{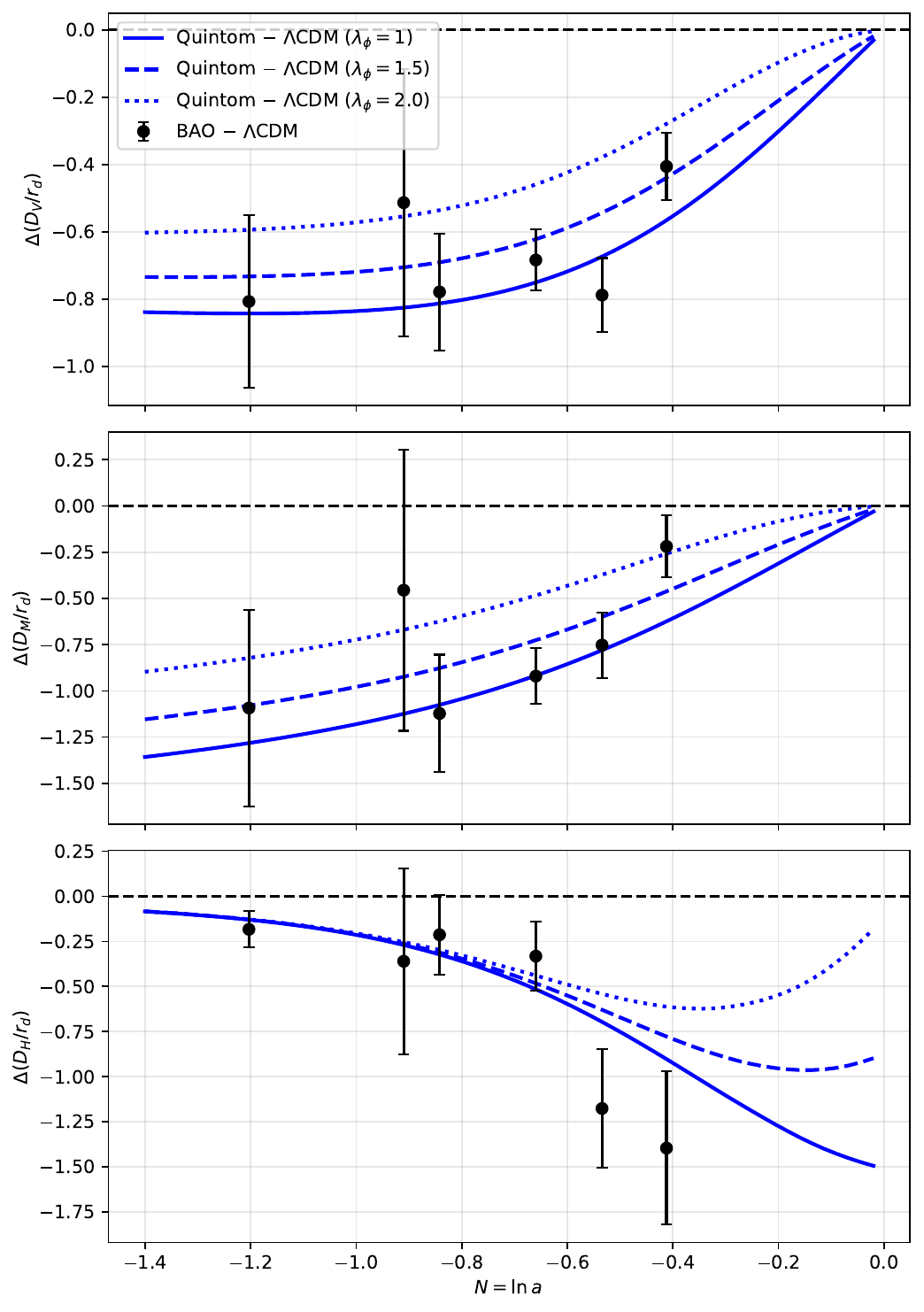}
            \caption{Plot of BAO distance differences  of the quintom model agaist the $\Lambda$CDM model for three different choices of the $\lambda_\phi$ parameter. The top panel shows the differences for $D_V / r_d$, the middle panel is for $D_M / r_d$ and the bottom panel is for $D_H/r_d$. The diffence with the observed data is shown in black points with correspoonding error bars.}
            \label{fig:BAO}
    \end{figure}

In Fig.~\ref{fig:BAO}, we present the differences in BAO distance measures predicted by the quintom model relative to the 
$\Lambda$CDM model, together with the corresponding differences computed from DESI DR2 observations (shown in black). The figure indicates that the quintom model yields a better fit to the BAO data than 
$\Lambda$CDM.

\subsection{State Finder Analysis}

To discriminate the model discussed here from other phenomenologically degenerate dark energy scenarios, we perform a statefinder analysis using the diagnostic pair $\{r, s\}$ constructed from the scale factor and its higher-order time derivatives \cite{statefinder_Chiba, statefinder_Sahni}. These parameters are defined as:

\begin{equation}\label{def:statefinder}
    \{r, s\} \equiv \left\{\frac{\dddot{a}}{aH^3}\ , \ \frac{r - 1}{3(q - \frac{1}{2})}\right\}\ .
\end{equation}

For an exponentially expanding universe, $a \propto e^t$, one finds, $q = -1$ and $r = 1$, which implies $s = 0$. This corresponds to the late time behaviour in the $\Lambda$CDM model. The evolution of these parameters is shown in Fig.~\ref{fig:rqs}. Since different dark energy models trace qualitatively distinct trajectories in the $(r-s)$ plane, the statefinder diagnostic provides a tool to distinguish among them. Using Friedmann, acceleration, and conservation equations, we express these parameters in terms of $w_i$, its time derivatives $\dot{w}_i$ \cite{statefinder_quintom}: 

\begin{equation}
\begin{split}
    q &= -\frac{\ddot{a}}{aH^2}=  \frac{1}{2}(1+3w_i \Omega_{i})\ ,\\
    r &= 1 + \frac{9}{2}\left(1 + w_i\right)\frac{\dot{p}_i}{\dot{\rho}_i} = 1 + \frac{9}{2}\left(w_i(1+w_i) - \frac{\dot{w_i}}{3H}\right)\ , \\
    s &=\frac{(1 + w_i)}{w_i} \frac{\dot{p}_i}{\dot{\rho}_i} = 1 + w_i -\frac{\dot{w}_i}{3Hw_i}\ .
\end{split}
\end{equation}

Focusing on the late time dynamics, and noting that the parameter $s$ is ill-defined for pressureless matter, we treat the total dark energy sector as a single effective fluid with equation of state $w_i \equiv w_{\rm DE}$. Including the radiation component, they can be written as:

\begin{equation}
\begin{split}
    q &= \frac{1}{2}(1+3w_{\rm DE}\Omega_{\rm DE} + \Omega_r)\ ,\\
    r &= 1 + \frac{9\Omega_{\rm DE}}{2}\left(w_{\rm DE}(1+w_{\rm DE}) - \frac{w'_{\rm DE}}{3}\right) + 2\Omega_r\ , \\
    s &= 1 + w_{\rm DE}\Omega_{\rm DE} -\frac{w'_{\rm DE}}{3w_{\rm DE}} + \frac{\Omega_r}{3}\ .
\end{split}
\end{equation}

The corresponding trajectories in the $r-s$ plane are displayed in Fig.~\ref{fig:statefindrqs}. The near linearity in the evolution in the past and more pronounced toward the future direction reflects the transition of the deceleration parameter from a matter dominated phase with $q \simeq 0.5$ to a dark-energy-dominated phase with $q \simeq -1$.

Finally, in Fig.~\ref{fig:thaw_freeze} we present the phase space diagram $w'_{\rm DE}$ vs $w_{\rm DE}$ following the classification introduced in \cite{Caldwell:2005tm}. The green (blue) region corresponds to the thawing (freezing) behaviour of the dark energy equation of state. While canonical quintessence models typically remain confined to one of these regions, quintom models allow for richer dynamics, exhibiting trajectories that go across thawing ($(1 + w_{\rm DE}) < w'_{\rm DE} < 3(1 + w_{\rm DE})$) and freezing ($3w_{\rm DE} (1 + w_{\rm DE}) < w'_{\rm DE} < 0.2w_{\rm DE}(1 + w_{\rm DE})$) regions at different cosmological epochs.

\section{Observational Data}\label{Observational Data}

In order to evaluate the performance of this model in relation to contemporary cosmological observations, we analyze several datasets using a Markov Chain Monte Carlo (MCMC) approach implemented with the \textit{COBAYA}\cite{Torrado:2020dgo} package. The background cosmological evolution described by Eq.~(\ref{eq:autonomous1}) is implemented in a separate Python module,
which is interfaced with \textit{COBAYA} as an external cosmological theory module for the MCMC analysis. In addition, we use the publicly available \textit{GetDist} package to visualize and analyze the resulting posterior distributions.

\subsection{Supernova Data}
Standard candles often include Type Ia supernovae due to their notably consistent absolute luminosity~\cite{reiss1998supernova,SupernovaSearchTeam:1998fmf}. In our current study, we used the Pantheon Plus compilation of SN-Ia datasets~\cite{Scolnic:2021amr, Riess_2022, Brout:2022vxf} in conjunction with the DES Year 5 data \cite{DES:2024jxu}. These collections are characterized by unique photometric systems and selection approaches, offering distance moduli $\mu$ across a span of redshifts.

\subsection{DESI BAO Data}

The visible baryonic matter density demonstrates repeating, periodic variations known as baryon acoustic oscillations. These oscillations are crucial standard rulers for accurate cosmological distance measurements. In this work, the 2025 BAO data from the Dark Energy Spectroscopic Instrument (DESI DR2), as cited in reference ~\cite{DESI:2025zgx, DESI:2024mwx}, have been utilized \footnote{The DESI DR2 data used in this analysis can be found in \href{https://github.com/CobayaSampler}{https://github.com/CobayaSampler}.}. BAO provides measurements of the effective distance parallel to the line of sight as follows, 

\begin{equation} 
\frac{D_H(z)}{r_d} =\frac{c r_d^{-1}}{H(z)} \ ,
\end{equation} 

and perpendicular to the line of sight as, 

\begin{equation} 
\frac{D_{M}(z)}{r_{d}}\equiv\frac{c}{r_{d}}\int_{0}^{z}\frac{d\tilde{z}}{H(\tilde{z})}=\frac{c}{H_{0}r_{d}}\int_{0}^{z}\frac{d\tilde{z}}{h(\tilde{z})}. 
\end{equation} 

The angle-averaged distance is calculated as, 

\begin{equation} 
\frac{D_V(z)}{r_d} =\left[\frac{c z r_d^{-3} d_L^2(z)}{H(z)(1+z)^2}\right]^{\frac{1}{3}}\ , 
\end{equation} 

where $d_L(z)$ represents the luminosity distance. 
% The effective redshift and the corresponding ratios $D_M / r_d$, $D_H / r_d$, and $D_V / r_d$ cited in this study are detailed in TABLE: 1 of reference ~\cite{DESI:2024mwx}.

\subsection{Compressed CMB Likelihood}

In addition to late time distance probes, we incorporate cosmic microwave background (CMB) information using a compressed likelihood that encodes early universe physics in a model independent manner. Following the approach described in Appendix~A of Ref.~\cite{DESI:2025zgx}, the
full CMB power spectrum information is compressed into a correlated Gaussian prior on the parameters $(\theta_\ast, \omega_b, \omega_{bc})$, where $\theta_\ast$ denotes the angular size of the sound horizon at recombination and $\omega_b \equiv \Omega_b h^2 $ and $\omega_{bc} \equiv (\Omega_b + \Omega_c) h^2$ are the physical baryon and total matter densities, respectively. These quantities are tightly constrained by the CMB and can be determined largely independently of
assumptions about late time cosmic evolution by marginalizing over effects such as CMB lensing and the late integrated Sachs--Wolfe effect \cite{Lemos:2023xhs}.

This compressed CMB likelihood effectively provides a high-redshift anchor for the background expansion history and captures most of the relevant constraining power of the full CMB likelihood for dark energy studies. In particular, it strongly constrains the matter density, thereby breaking degeneracies present in low-redshift probes such as BAO. As demonstrated in Ref.~\cite{DESI:2025zgx}, constraints on evolving dark energy models obtained using this compressed likelihood are highly consistent with those derived from the full CMB likelihood,
indicating that the dominant CMB information relevant for background cosmology is retained. We therefore adopt this compressed CMB prior as a robust and computationally efficient alternative to fitting the full CMB power spectra.

We have examined the following three combinations of datasets.

\begin{enumerate}
    \item Set 1: Pantheon Plus + CMB + DESI DR2,
    \item  Set 2: DES Y5 + CMB + DESI DR2,
    \item Set 3: CMB + DESI DR2.
\end{enumerate}

\begin{figure*}[!hbt]
            \centering
            \includegraphics[width=\textwidth]{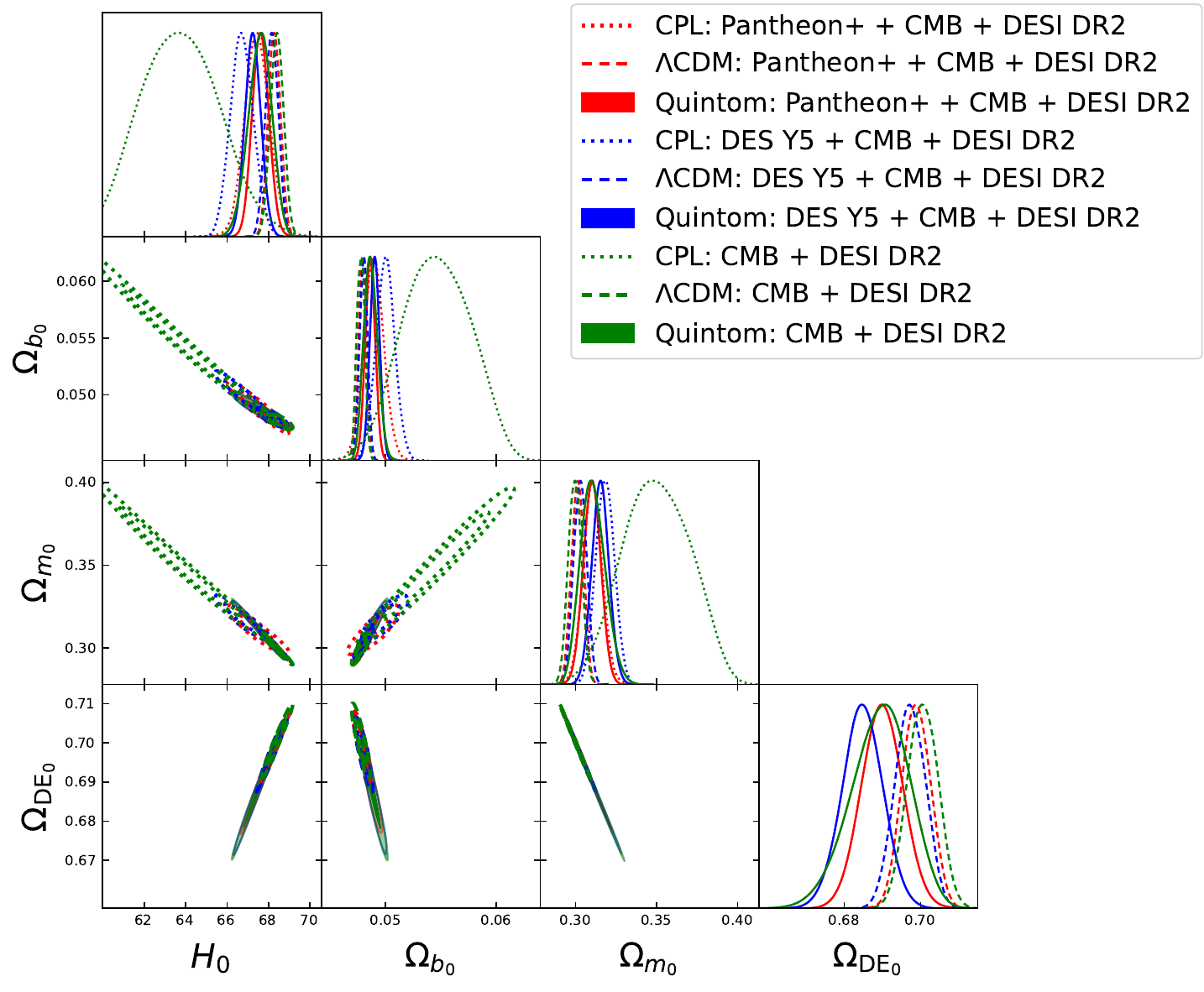}
            \caption{1D and 2D triangular plots of the posterior distributions of cosmological parameters for the model, $w_{0}w_{a}$ and $\Lambda$CDM model (dotted) are shown for comparison.}
            \label{fig:Final_Quintom_vs_lcdm}
\end{figure*}

    \begin{figure}[!hbt]
            \centering
            \includegraphics[width=\columnwidth]{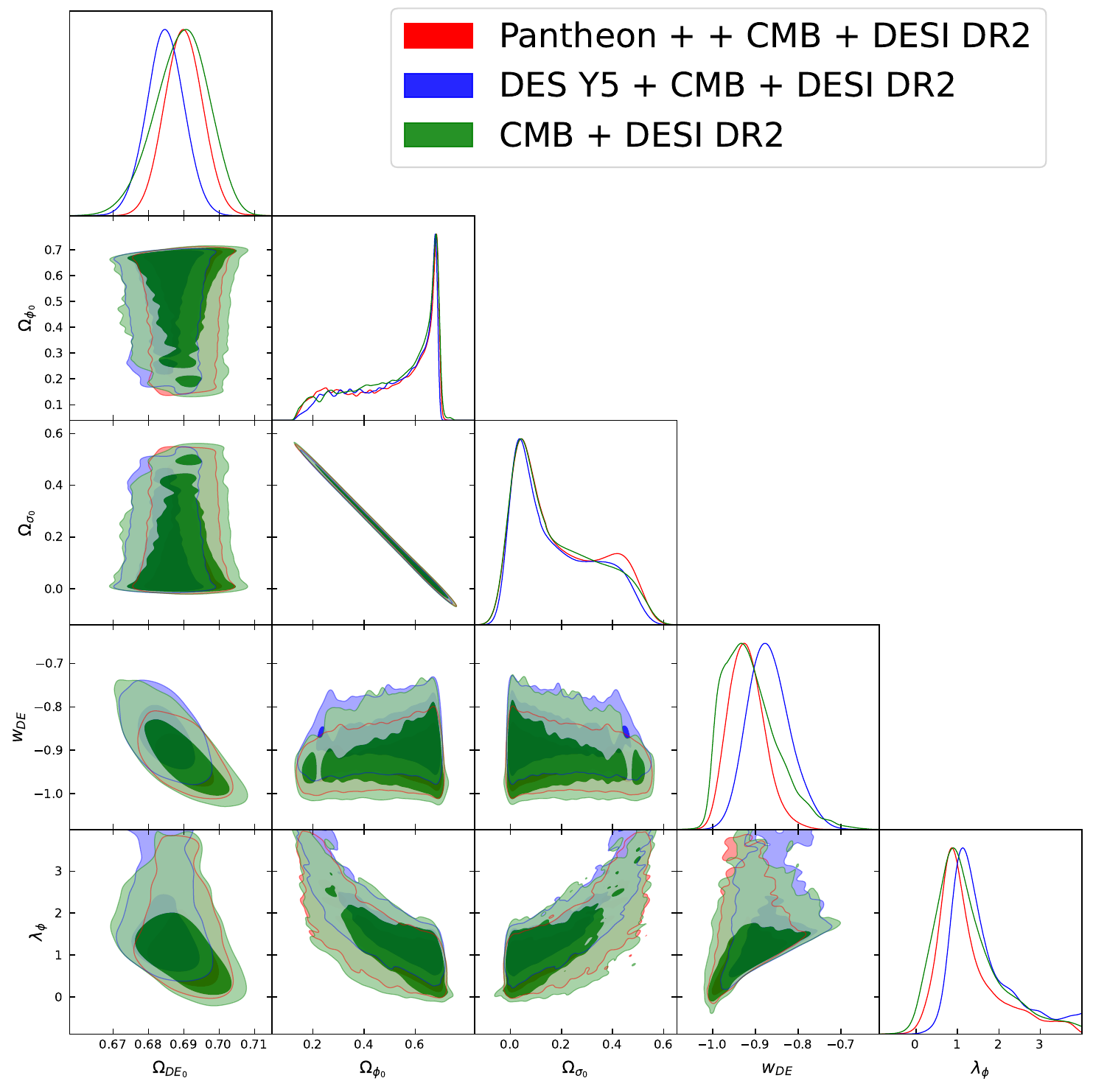}
            \caption{Triangular plots of the derived cosmological parameters and model parameters for the model.}
            \label{fig:Final_All_Quintom}
    \end{figure}

\begin{figure}[!hbt]
            \centering
            \includegraphics[width=\columnwidth]{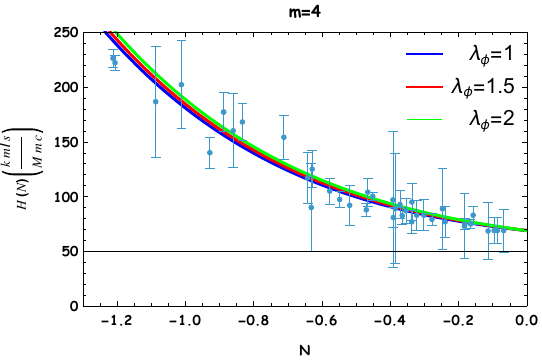}
            \caption{Evolution of Hubble parameter with observed values.}
            \label{fig:Hubble}
    \end{figure}

\begin{table*}[ht]
\centering
\resizebox{\textwidth}{!}{%
\begin{tabular}{lccccccccc}
\toprule
Parameters & \multicolumn{3}{c}{Set 1: Pantheon Plus+CMB+DESI DR2} & \multicolumn{3}{c}{Set 2: DES Y5+CMB+DESI DR2} & \multicolumn{3}{c}{Set 3: CMB+DESI DR2} \\
\cmidrule(lr){2-4}\cmidrule(lr){5-7}\cmidrule(lr){8-10}
& $\Lambda$CDM & Quintom & $w_{0}w_{a}$  & $\Lambda$CDM & Quintom & $w_{0}w_{a}$  & $\Lambda$CDM & Quintom & $w_{0}w_{a}$  \\
\midrule
$H_0$
& $68.28\pm 0.29$
& $67.64\pm 0.40$
& $67.46\pm 0.61$
& $68.16\pm 0.31$
& $67.25\pm 0.39$
& $66.72\pm 0.55$
& $68.41\pm 0.31$
& $67.57^{+0.60}_{-0.54}$
& $63.8^{+1.70}_{-2.10}$ \\

$\Omega_{b0}$
& $0.04789\pm 0.00032$
& $0.04859\pm 0.00045$
& $0.04893\pm 0.00091$
& $0.04801\pm 0.00035$
& $0.04902\pm 0.00045$
& $0.04997\pm 0.00085$
& $0.04776\pm 0.00035$
& $0.04867^{+0.00059}_{-0.00068}$
& $0.0546\pm 0.0030$ \\

$\Omega_{m0}$
& $0.3011\pm 0.0037$
& $0.3098\pm 0.0053$
& $0.3105\pm 0.0059$
& $0.3028\pm 0.004$
& $0.3153\pm 0.0053$
& $0.3181\pm 0.0056$
& $0.2995\pm 0.004$
& $0.3109^{+0.0071}_{-0.0084}$
& $0.350\pm 0.021$ \\

$\Omega_{\rm DE}$
& $0.6989\pm 0.0037$
& $0.6902\pm 0.0053$
& $0.6895\pm 0.0059$
& $0.6972\pm 0.004$
& $0.6847\pm 0.0053$
& $0.6819\pm 0.0056$
& $0.7005\pm 0.004$
& $0.6891^{+0.0084}_{-0.0071}$
& $0.650\pm 0.021$ \\

$\Omega_{\phi 0}$
& $-$
& $0.50^{+0.20}_{-0.16}$
& $-$
& $-$
& $0.517^{+0.18}_{-0.071}$
& $-$
& $-$
& $0.50^{+0.20}_{-0.29}$
& $-$ \\

$\Omega_{\sigma}$
& $-$
& $0.19^{+0.12}_{-0.22}$
& $-$
& $-$
& $0.168^{+0.089}_{-0.19}$
& $-$
& $-$
& $0.19^{+0.11}_{-0.21}$
& $-$ \\

$w_{\rm DE}$ 
& $-1$
& $-0.920^{+0.034}_{-0.045}$
& $-0.852\pm 0.055$
& $-1$
& $-0.867^{+0.042}_{-0.058}$
& $-0.766\pm 0.058$
& $-1$
& $-0.909^{+0.034}_{-0.087}$
& $-0.45\pm 0.20$ \\

$\lambda_{\phi}$
& $-$
& $1.3072\pm 0.8062$
& $-$
& $-$
& $1.5676\pm 0.7580$
& $-$
& $-$
& $1.3198\pm 0.855$
& $-$ \\

$\chi_{min}^{2}$ 
& $1419.93$
& $1410.86$
& $1412.869$
& $1663.82$
& $1646.01$
& $1646.292$
& $13.89$
& $7.88$
& $7.091$\\

$\Delta \chi_{min}^{2}$ 
& $0$
& $-9.07$
& $-7.061$
& $0$
& $-17.81$
& $-17.528$
& $0$
& $-6.02$
& $-6.799$\\

$AIC$
& $1427.93$
& $1426.86$
& $1422.869$
& $1671.82$
& $1662.01$
& $1656.292$
& $21.89$
& $23.88$
& $17.091$\\

$\Delta AIC$
& $0$
& $-1.07$
& $-5.061$
& $0$
& $-9.81$
& $-15.528$
& $0$
& $1.99$
& $-4.799$\\

$|\ln B_{i \Lambda}|$
& $0$
& $1.969$
& $2.749$
& $0$
& $3.3$
& $2.5$
& $0$
& $2.671$
& $1.079$\\

\bottomrule
\end{tabular}
}
\caption{Mean values and 68\% confidence intervals for the cosmological parameters, along with $\Delta\chi_{\text{min}}^2$, $\Delta \mathrm{AIC}$ and MCEvidence analyses, are presented for all of $\Lambda$CDM, $w_{0}w_{a}$ and the quintom model.}
\label{tab:bestfit}
\end{table*}

\section{Constraints from Data Analysis}\label{Constraints from Data Analysis}
In this section, the observational bounds are provided for the quintom model together with the $\Lambda$CDM model and $w_{0}w_{a}$ model for comparison using the combined set of three cosmological data samples mentioned previously. The present-day matter density parameter is defined as $\Omega_m = \Omega_b + \Omega_c$, where $\Omega_b$ and $\Omega_c$ denote the baryonic and cold dark matter components, respectively. The parameter estimation is performed by imposing flat (uniform) priors on the primary cosmological quantities, namely $H_0:[60,80]~\mathrm{km\,s^{-1}\,Mpc^{-1}}$, $\Omega_{b0}:[0.01,0.1]$ and $\Omega_{m0}:[0.1,0.5]$. For the quintom sector, additional parameters associated with the dynamical degrees of freedom are allowed to vary within the ranges $x_{\phi 0}\in[0,0.4]$, $x_{\sigma 0}\in[-0.02,0.25]$, $y_{\sigma 0}\in[-0.75,0.75]$, $\lambda_{\phi}\in[-2,4]$, and $\lambda_{\sigma 0}\in[-2,2]$, where the subscript ``0'' denotes present-day values. These parameters encode the initial conditions of the scalar field dynamics as well as the slope of the quintessence potential.

Table~\ref{tab:bestfit} summarizes the best-fit (mean) values together with the corresponding $68\%$ confidence-level limits for both the cosmological and model parameters obtained in our study. These constraints are derived using the three different combinations of observational data sets introduced in the previous section. For reference, we also display the results obtained for the  $\Lambda$CDM and $w_{0}w_{a}$ scenario.

The marginalized one and two-dimensional posterior distributions of the background cosmological parameters $H_0$, $\Omega_{b0}$, $\Omega_{m0}$, and $\Omega_{\rm DE0}$ are displayed in Fig.~\ref{fig:Final_Quintom_vs_lcdm}. The corresponding constraints on the dark-energy sector parameters, including $\Omega_{\phi 0}$, $\Omega_{\sigma 0}$, $w_{\rm DE}$, and the model parameter $\lambda_{\phi}$, are shown in Fig.~\ref{fig:Final_All_Quintom}. In these figures, the results obtained from data Set~1, Set~2, and Set~3 are represented by red, green, and blue contours, respectively. The $\Lambda$CDM and $w_{0}w_{a}$ constraints are indicated by dashed and dotted curves, while solid curves correspond to the quintom model.

Across all three data combinations, the inferred values of the Hubble constant in both the quintom and $w_{0}w_{a}$ scenarios are slightly lower than those obtained in the $\Lambda$CDM case, although the shifts remain within the $1\sigma$ confidence level and therefore do not indicate a statistically significant deviation. The matter density parameter $\Omega_{m0}$ is systematically higher in the dynamical dark energy models, particularly in the $w_{0}w_{a}$ parametrization, with a corresponding reduction in the dark energy density parameter $\Omega_{\rm DE0}$. This trend reflects the fact that the effective dark energy equation of state in these dynamical dark energy models deviates from $w=-1$ towards the $w>-1$ at current epoch. In particular, the $w_{0}w_{a}$ model exhibits the largest deviation from $w=-1$, while the quintom model remains closer to the cosmological constant.

The decomposition of the dark energy sector reveals that the present cosmic acceleration is predominantly driven by the canonical scalar field, with $\Omega_{\phi 0}\simeq 0.5$, while the phantom component contributes subdominantly, with $\Omega_{\sigma 0}\sim 0.17$ - $0.19$ depending on the data set. The total dark energy equation-of-state parameter is constrained to lie in the range $-0.92 \lesssim w_{\rm DE} \lesssim -0.87$, indicating a clear departure from a pure cosmological constant and favouring a dynamical dark energy behaviour.

The relative statistical performance of the quintom model with respect to $\Lambda$CDM is quantified using the minimum chi-square value $\chi^2_{\rm min}$, its difference $\Delta\chi^2_{\rm min}$, and the Akaike Information Criterion (AIC). As shown in Table~\ref{tab:bestfit}, the quintom model yields lower $\chi^2_{\rm min}$ values for all three data combinations, with the most significant improvement obtained when DES~Y5 data are included. This indicates that the extended model provides a better fit to the data at the level of the likelihood.

The difference in the Akaike Information Criterion is defined as
\begin{equation}\label{eq:AIC}
\Delta \mathrm{AIC}
= \chi^{2}_{\rm min,\mathcal{M}} - \chi^{2}_{\rm min,\Lambda{\rm CDM}}
+ 2\left(N_{\mathcal{M}} - N_{\Lambda{\rm CDM}}\right),
\end{equation}
where $\mathcal{M}$ denotes the model under consideration and $N_{\mathcal{M}}$ represents the number of free parameters sampled in the MCMC analysis. A negative value of $\Delta\mathrm{AIC}$ indicates a preference for the model $\mathcal{M}$ after accounting for its increased complexity.

According to the AIC results reported in Table~\ref{tab:bestfit}, the quintom model is mildly favored over $\Lambda$CDM for the Pantheon Plus+CMB+DESI DR2 data set and strongly favored when DES~Y5 data are included. In contrast, for the CMB+DESI DR2 combination, the AIC slightly favors $\Lambda$CDM. From the AIC analysis, the $w_0w_a$ model shows a stronger preference compared to the quintom model over $\Lambda$CDM. This result is expected, since the quintom model incorporates more free parameters than the $w_0w_a$ model.

In addition to the  model comparison based on $\chi^2_{\rm min}$ and AIC, we also assess the relative performance of the quintom scenario using Bayesian evidence. The Bayesian evidence $\mathcal{Z}$ corresponds to the likelihood integrated over the full parameter space weighted by the prior distributions. In practice, we evaluate the logarithmic Bayes factor between the quintom model and the $\Lambda$CDM scenario using the \texttt{MCEvidence}\cite{Heavens:2017afc} algorithm applied to the MCMC chains.

The resulting values of $\ln B_{i \Lambda} \equiv \ln(\mathcal{Z}_{\rm Model}/\mathcal{Z}_{\Lambda{\rm CDM}})$ are reported in Table~\ref{tab:bestfit}. According to the Jeffreys scale, values of $|\ln B_{i \Lambda} | \lesssim 1$ indicate inconclusive evidence, values in the range $1 \lesssim |\ln B_{i \Lambda} | \lesssim 2.5$ correspond to weak evidence, values in the range $2.5 \lesssim |\ln B_{i \Lambda} | \lesssim 5$ correspond to moderate evidence, values in the range $5 \lesssim |\ln B_{i \Lambda} | \lesssim 10$ correspond to strong evidence, while values exceeding $|\ln B_{i \Lambda} | \gtrsim 10$ is interpreted as very strong evidence.

For the datasets including supernova observations (Set~1 and Set~2), the Bayesian evidence analysis indicates a preference for dynamical dark energy models over the $\Lambda$CDM scenario. For Set~1 (Pantheon Plus+CMB+DESI DR2), the Bayes factors are $\ln B_{Q\Lambda}=1.969$ for the quintom model and $\ln B_{w_{0}w_{a}\Lambda}=2.749$ for the $w_{0}w_{a}$ parametrization. According to the Jeffreys scale, these values correspond to weak evidence in favor of the quintom model and moderate evidence in favor of the $w_{0}w_{a}$ model relative to $\Lambda$CDM.

Similarly, for Set~2 (DES Y5+CMB+DESI DR2), the Bayes factors $\ln B_{Q\Lambda}=3.3$ and $\ln B_{w_{0}w_{a}\Lambda}=2.5$ indicate moderate evidence supporting both dynamical dark energy scenarios over $\Lambda$CDM, with a slightly stronger preference for the quintom model in this dataset. 

For Set~3 (CMB+DESI DR2), the Bayesian evidence also favors the dynamical dark energy models over $\Lambda$CDM. In this case the Bayes factors are $\ln B_{Q\Lambda}=2.671$ for the quintom model and $\ln B_{w_{0}w_{a}\Lambda}=1.079$ for the $w_{0}w_{a}$ model.  

These findings indicate that although the dynamical models achieve a better fit to the data evidenced by the reductions in the minimum chi-square values ($\Delta \chi_{\rm min}^{2} = -9.07$ for Set~1, $\Delta \chi_{\rm min}^{2} = -17.81$ for Set~2 and $\Delta \chi_{\rm min}^{2} = -6.02$ for Set~3 in the quintom scenario)—their expanded parameter space weakens their statistical advantage in a Bayesian framework. In other words, despite the improved $\chi_{\rm min}^{2}$, the present datasets do not provide decisive Bayesian evidence in favor of the quintom model over $\Lambda$CDM. This outcome is characteristic of extended dark energy models with additional dynamical degrees of freedom, where the gain in goodness of fit is counterbalanced by the Bayesian penalty for increased model complexity.

\section{Conclusion}\label{Conclusion}

In this work, we have presented a comprehensive study of a quintom dark energy model: a quintessence field with an exponential potential and a phantom degree of freedom governed by an inverse power-law potential. By introducing suitable dimensionless variables, the background equations were cast into an autonomous system, allowing for a systematic classification of the fixed points and their stability properties. The phase space analysis revealed a rich dynamical structure, including matter dominated saddle points, kinetically dominated early time solutions, and mixed scalar field configurations that can support accelerated expansion. Among all critical points, we identified a unique late time attractor corresponding to a phantom potential dominated de Sitter phase. Although this attractor is non-hyperbolic, numerical phase space analysis confirms its stability within the physically relevant region, indicating that the model generically evolves toward phantom dominated accelerated expansion at late times.

An important outcome of this analysis is that the cosmological evolution does not proceed directly toward the asymptotic phantom dominated attractor. Instead, the universe undergoes an intermediate epoch in which the canonical scalar field dominates the dark energy sector including the current epoch, while the phantom component remains subdominant. The effective dark energy equation of state remains close to $w\simeq-1$ throughout the matter dominated era and subsequently evolves toward less negative values at low redshift, exhibiting a thawing-like behaviour. Both the fields becomes dynamically relevant only at later times, eventually driving the universe toward the complete phantom domination.

In this model, phantom crossing is achieved by setting the initial conditions within the phantom region during the radiation dominated epoch. A notable feature of the phantom crossing obtained here is that the crossing of the $w\simeq-1$ barrier occurs gradually and asymptotically, rather than abruptly. Using the dynamical systems framework, we demonstrate that this behaviour of phantom crossing is generic for this class of quintom models in which the scalar fields are not coupled at the level of the potential.

We confronted the model with current cosmo logical observations using several combinations of low and high-redshift data, including Type Ia supernovae, cosmic microwave background distance priors and baryon acoustic oscillations. The results show that the quintom scenario provides a fit to the data that is at least comparable to that of the standard $\Lambda$CDM model and, in some data combinations, leads to a modest improvement in the minimum $\chi^2$. Infor mation criteria analyses indicate that the statistical preference for the quintom model is dataset depen dent, with mild to moderate support.

\section*{Acknowledgement}
PHP acknowledges financial support from \textit{Fundação de Amparo à pesquisa e Inovação do Espírito Santo} (FAPES, Brazil) and is grateful to the Departamento de F\'isica Te\'orica, UERJ and the ITA, University of Oslo, for their hospitality. Authors are also thanful to L. Arturo Ure\~{n}a-L\'{o}pez for helping with the implimentation of compresed CMB likelihood in \textit{COBAYA}.

%\appendix

%\section{Theory}
%\input{models}

%\clearpage
\bibliographystyle{unsrt}
\bibliography{sample}

\end{document}